\documentclass[review]{elsarticle}

\usepackage{lineno,hyperref}
\modulolinenumbers[5]
\usepackage{geometry}
\usepackage[table,xcdraw]{xcolor}
\usepackage{caption}
\usepackage{subcaption}
\usepackage{array}

\begin{document}

\begin{frontmatter}

\title{Predicting Ovarian Cancer Treatment Response in Histopathology using Hierarchical Vision Transformers and Multiple Instance Learning}


\author[cistib]{Jack Breen}\corref{mycorrespondingauthor}\cortext[mycorrespondingauthor]{Corresponding author. Email address: scjjb@leeds.ac.uk (Jack Breen)}

\author[limr]{Katie Allen}

\author[limr]{Kieran Zucker}

\author[limr]{Geoff Hall}

\author[cistib]{Nishant Ravikumar\corref{cor1}}

\author[limr]{Nicolas M Orsi\corref{cor1}}

\address[cistib]{Centre for Computational Imaging and Simulation Technologies in Biomedicine (CISTIB), School of Computing, University of Leeds, UK}
\address[limr]{Leeds Institute of Medical Research at St James's, School of Medicine, University of Leeds, UK}
\cortext[cor1]{Indicates joint last authors}

\begin{abstract}
For many patients, current ovarian cancer treatments offer limited clinical benefit. For some therapies, it is not possible to predict patients' responses, potentially exposing them to the adverse effects of treatment without any therapeutic benefit. As part of the \emph{automated prediction of treatment effectiveness in ovarian cancer using histopathological images} (ATEC23) challenge, we evaluated the effectiveness of deep learning to predict whether a course of treatment including the antiangiogenic drug \emph{bevacizumab} could contribute to remission or prevent disease progression for at least 6 months in a set of 282 histopathology whole slide images (WSIs) from 78 ovarian cancer patients. Our approach used a pretrained Hierarchical Image Pyramid Transformer (HIPT) to extract region-level features and an attention-based multiple instance learning (ABMIL) model to aggregate features and classify whole slides. The optimal HIPT-ABMIL model had an internal balanced accuracy of 60.2\%$\pm$2.9\% and an AUC of 0.646$\pm$0.033. Histopathology-specific model pretraining was found to be beneficial to classification performance, though hierarchical transformers were not, with a ResNet feature extractor achieving similar performance. Due to the dataset being small and highly heterogeneous, performance was variable across 5-fold cross-validation folds, and there were some extreme differences between validation and test set performance within folds. The model did not generalise well to tissue microarrays, with accuracy worse than random chance. It is not yet clear whether ovarian cancer WSIs contain information that can be used to accurately predict treatment response, with further validation using larger, higher-quality datasets required. 
\end{abstract}


\journal{ATEC23 Challenge}

\end{frontmatter}


\section{Introduction}
Ovarian cancer is the most lethal gynaecological malignancy worldwide, with almost 314,000 new cases diagnosed annually resulting in over 205,000 deaths  \cite{Sung2021}. Treatment options are guided by the stage, grade, and morphological subtype of ovarian cancer, and can often involve surgery, chemotherapy, and increasingly, immunotherapy. However, response to therapy can vary significantly, and the underlying causes are not well understood despite significant progress in defined subgroups, such as homologous recombination deficient tumours \cite{Miller2022}. Due to this knowledge gap, some patients may be exposed to the adverse effects of a given therapy without deriving any clinical benefit. The \emph{automated prediction of treatment effectiveness in ovarian cancer using histopathological images} (ATEC23) challenge aims to identify non-responders using pre-treatment histopathology whole slide images (WSIs) alone. 

Using artificial intelligence to make prognostic predictions from ovarian cancer histopathology images is a relatively new area of research, with few previous studies predicting treatment response \cite{Breen2023review}. 
Studies reporting higher accuracy in this particular area have used  immunohistochemistry (IHC) panels \cite{Wang2023}, with performance being poorer in studies using haematoxylin and eosin (H\&E)-stained tissue \cite{Yaar2020, Wang2022b}. A prediction model using H\&E WSIs alone would offer greater clinical benefit given that this staining method is routine in all histopathological diagnostic interpretation of ovarian cancer specimens. Instead, dependence on IHC staining would add financial and time burdens to the diagnostic pathway.

An H\&E baseline model was developed by the ATEC23 challenge organisers \cite{Wang2022b}, in which a hierarchical attention approach was used to segment the most relevant tissue. Attention-based multiple instance learning (ABMIL) \cite{Ilse2018} was then applied to this segmented tissue to classify WSIs. The reported results from 5-fold cross-validation presented an accuracy of 88.2\% and an F1 score of 0.917, although the reported accuracy on an independent test set was no greater than random guessing.

None of the previous ovarian cancer treatment response studies have employed methods that capture spatial relationships within WSIs, such as vision transformers \cite{Dosovitskiy2020} or graph networks \cite{Li2018}. Such methods are likely to be beneficial as there are established correlations between patient prognosis and the spatial arrangement of cellular structures visible in WSIs, with tumour-infiltrating lymphocytes being associated with survival in some ovarian cancer subtypes \cite{Goode2017}. In this study, we combined vision transformers with ABMIL to classify whether patients will respond to a specific course of bevacizumab-based therapy from histopathology WSIs alone, as defined by measurable recurrence/progression within 6 months of treatment.

\section{Methods}

\begin{figure}[h]
    \centering
    \includegraphics[width=0.9\textwidth]{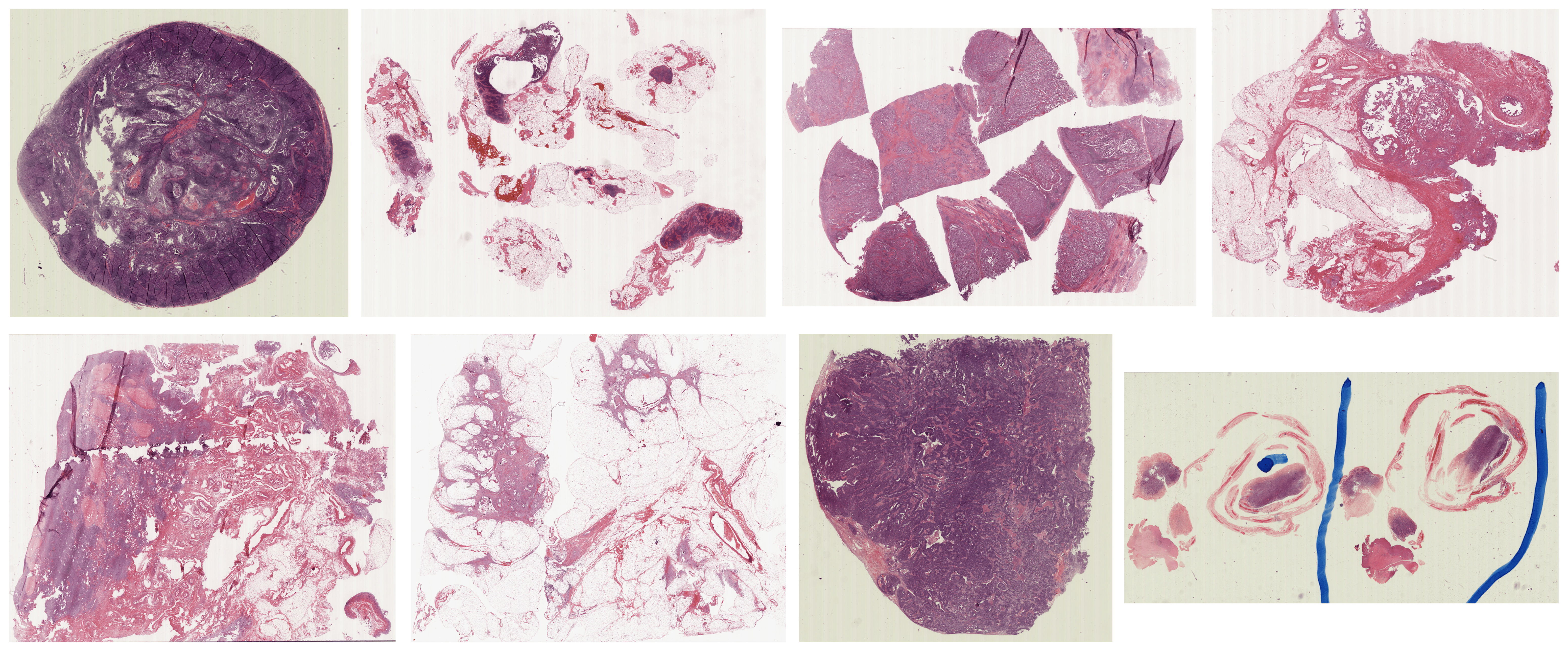}
    \caption{Eight whole slide images from the ATEC23 challenge training set}
    \label{fig:thumbnails}
\end{figure}

The challenge training data \cite{Wang2022} comprised 288 H\&E-stained tissue section WSIs from 78 tubo-ovarian and primary peritoneal cancer patients, of which 53 were determined to have an \emph{effective} response to treatment, and 25 were determined to have an \emph{invalid} response to treatment. We used 282 WSIs from 78 patients due to two WSIs being inaccessible, two being duplicated, and two being erroneously excluded. All patients received debulking surgery, chemotherapy, and bevacizumab therapy, with treatment classified as \emph{effective} if CA-125 levels fell and there was no tumour progression/recurrence found in CT/PET images within 6 months of treatment. All samples were originally collected from a single data centre and scanned using a single Leica AT Turbo scanner at 20x magnification, with these scans made available in the ATEC23 challenge. Patients had a range of morphological subtype diagnoses, including high-grade serous (n=58), clear cell (n=7), unclassified (n=7), endometrioid (n=4), and mucinous carcinomas (n=2). The slides in the dataset were highly heterogeneous (Figure \ref{fig:thumbnails}). Samples appeared to include a combination of adnexal, omental, and lymph node tissue, with some slides having differing colour profiles and artifacts, such as pen markings. An independent challenge test set was collected at the same data centre, consisting of 180 H\&E-stained tissue microarray (TMA) single core images from patients with high-grade serous ovarian carcinoma.

\begin{figure}[h]
    \centering
    \includegraphics[width=0.9\textwidth]{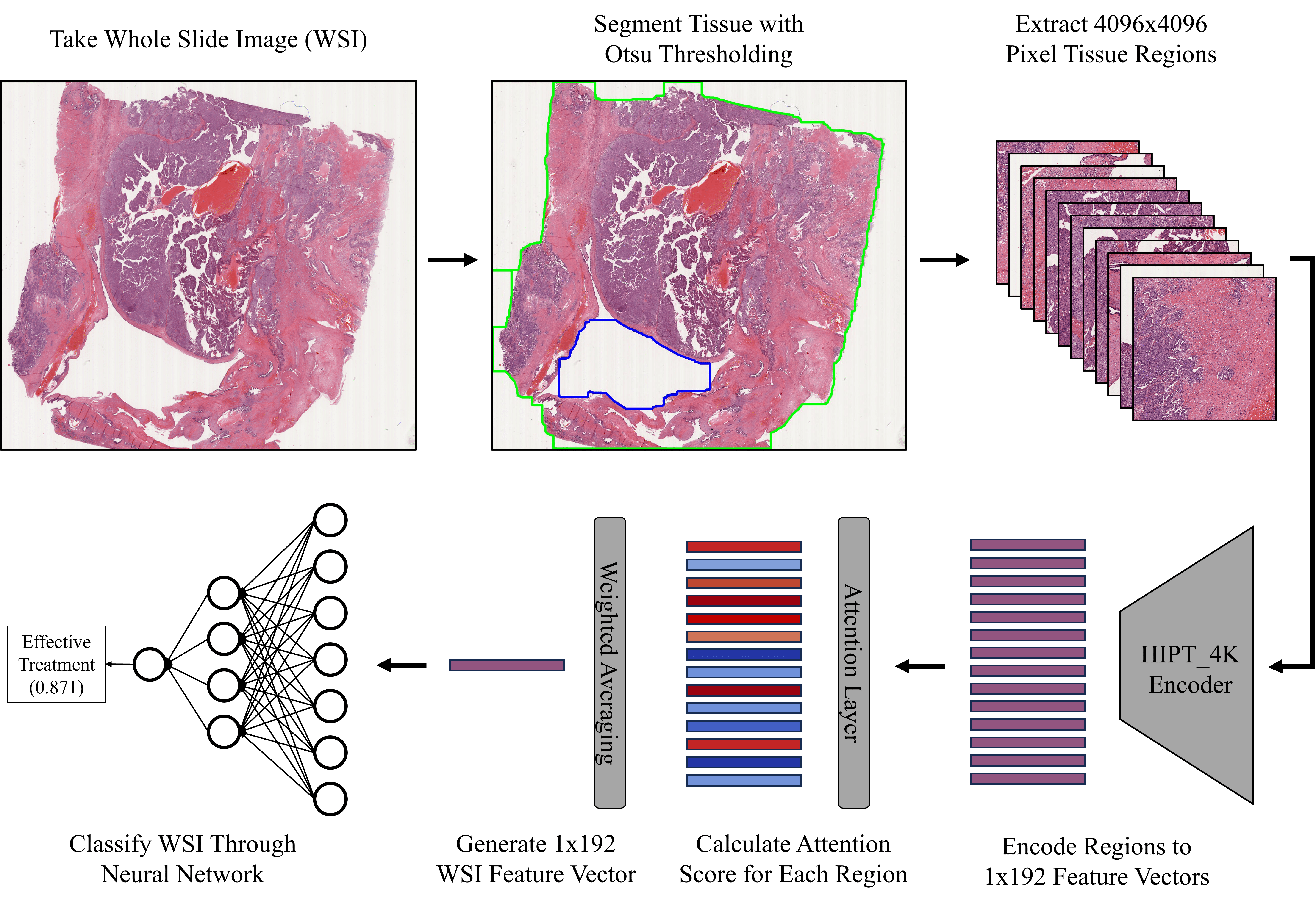}
    \caption{HIPT-AMBIL whole slide image classification pipeline}
    \label{fig:HIPT-ABMIL}
\end{figure}

Our HIPT-ABMIL classification approach, shown in Figure \ref{fig:HIPT-ABMIL}, used ABMIL \cite{Ilse2018} to classify WSIs based on region-level (4096x4096 pixel) features encoded through a two-stage vision transformer \cite{Chen2022}. Before modelling, we used Otsu thresholding to segment tissue, then extracted 4096x4096 non-overlapping tissue regions for modelling. On average, the tissue patching procedure generated 91 regions per slide (range of 13 to 166).

We extracted features from each tissue region using the two-stage Hierarchical Image Pyramid Transformer (HIPT\_4K) \cite{Chen2022}. This approach first uses a vision transformer \cite{Dosovitskiy2020} to aggregate cell-level information (16x16 pixels) to patch-level (256x256), and subsequently uses a second vision transformer to aggregate patch-level information to region-level (4096x4096). This feature extractor was pretrained using over ten thousand total histopathology slides from 33 cancer types using the self-supervised method \emph{DINO} \cite{Caron2021}. We trained the ABMIL network using these region embeddings to classify WSIs, termed HIPT-ABMIL. We also compared three other approaches - HIPT-CLAM, ResNet-ABMIL, and HistoResNet-ABMIL. \emph{Clustering-constrained-attention multiple instance learning} (CLAM) \cite{Lu2021} adds a clustering task to model training to encourage the model to learn discriminative features. ResNet-ABMIL is a standard approach in histopathology in which 256x256 pixel patches are encoded with an ImageNet-pretrained ResNet50 encoder, before ABMIL is used to aggregate patch embeddings and classify WSIs. HistoResNet-ABMIL is the same model but with features extracted through a ResNet18 encoder which was pretrained on a collection of 57 histopathology datasets \cite{Ciga2022} using the self-supervised technique \emph{SimCLR} \cite{Chen2020}. Due to the smaller patch size in the ResNet approaches, there were more patches per slide, with an average of 20214 (range of 2043 to 38828). 

Our training procedure included multiple approaches to reduce overfitting. We randomly selected a different subset of the available regions in each WSI every epoch as an efficient data augmentation technique, building upon our previous work in which we showed that a similar approach could improve slide evaluation efficiency without drastically reducing classification accuracy \cite{Breen2023spie}. We included a dropout layer in the classification network, which randomly dropped a proportion of the model weights during each training epoch. We also used weight decay in the Adam optimizer, which imposed an L2 regularisation term to disincentivise learning large weights in the network. 

We trained our models using a cross-entropy loss and an Adam optimiser. As shown in Table \ref{tab:tuning}, we tuned hyperparameters across 5-fold cross-validation experiments, using a grid search strategy for five hyperparameters. The parameters were the learning rate, dropout probability, L2 regularisation weight, attention layer size, and number of patches per slide for training. The attention layer size hyperparameter controlled the dimension of the attention layer, and the subsequent hidden layer in the classification network had a dimension half this size. One extra hyperparameter, B, was tuned for the CLAM model, which controlled the number of regions which were clustered in feature space during training. Each tuning configuration was repeated three times and the average loss was taken to account for random variations. Multiple stages of hyperparameter grid tuning were used, with earlier runs covering a wider range of parameters and influencing the hyperparameter options available in later stages. Each model was evaluated with over 500 total hyperparameter configurations.  

We selected the hyperparameters which minimised the average validation loss across the 5-fold cross-validation to train the final model. Internal performance was measured on the cross-validation test sets, and the same hyperparameters were used to train a 4-fold ensemble model with 75\%-25\% train-val splits, with the mean predictions for the external TMA images submitted to the ATEC23 challenge. Due to the relatively small size of test set images, each one was represented as a single 4096x4096 region. Our PyTorch-based code, an extension of the CLAM pipeline \cite{Lu2021}, is available at \mbox{\url{https://github.com/scjjb/HIPT_ABMIL_ATEC23}}, alongside further details of the hyperparameter tuning. Experiments used an NVIDIA DGX A100 server with 8 NVIDIA A100 GPUs and 256 AMD EPYC 7742 CPUs @3.4GHz.

\begin{table}[h]
\begin{center}
\resizebox{0.98\textwidth}{!}{%
\begin{tabular}{|
>{\columncolor[HTML]{707070}}c |c|c|c|c|c|}
\hline
{\color[HTML]{FFFFFF} Hyperparameter}    & \cellcolor[HTML]{707070}{\color[HTML]{FFFFFF} Function}                                                                  & \cellcolor[HTML]{707070}{\color[HTML]{FFFFFF} Initial Tuning Options} & \cellcolor[HTML]{707070}{\color[HTML]{FFFFFF} Second Tuning Options} & \cellcolor[HTML]{707070}{\color[HTML]{FFFFFF} Third Tuning Options} & \cellcolor[HTML]{707070}{\color[HTML]{FFFFFF} Final Selection} \\ \hline
{\color[HTML]{FFFFFF} Learning Rate}     & \begin{tabular}[c]{@{}c@{}}Sets the rate of change of model parameters\\ trained using the Adam optimiser\end{tabular}   & 1e-3, 1e-4, 1e-5                                                      & 1e-3, 5e-4, 1e-4                                                     &  1e-3, 5e-4 &  1e-3                                                         \\ \hline
{\color[HTML]{FFFFFF} Dropout}          & \begin{tabular}[c]{@{}c@{}}Sets the proportion of model weights to\\ drop in each training iteration\end{tabular}        & 0.25, 0.5, 0.75                                                    & 0.6, 0.75, 0.9                                                     &    0.8, 0.85, 0.9, 0.95       &  0.85                                               \\ \hline
{\color[HTML]{FFFFFF} Regularisation}    & \begin{tabular}[c]{@{}c@{}}Sets the level of weight decay\\ in the Adam optimiser\end{tabular}                           & 1e-2, 1e-3, 1e-4                                                     & 1e-1, 1e-2, 1e-3                                                     &         1e-0, 5e-1, 1e-1, 5e-2      &  5e-1                                           \\ \hline
{\color[HTML]{FFFFFF} \begin{tabular}[c]{@{}c@{}}Attention\\Layer Size\end{tabular} }        & \begin{tabular}[c]{@{}c@{}}Sets the dimension of the attention layer, with the \\ subsequent hidden layer size set to half of this\end{tabular}                  & 64, 32, 16                                                    & 32, 16, 8                                                       &    32, 16    & 16                                                \\ \hline
{\color[HTML]{FFFFFF} Patches per Slide} & \begin{tabular}[c]{@{}c@{}}Set the number of patches randomly\\ selected from each slide per training epoch\end{tabular} & 25, 50, 75                                                       & 25, 50, 75                                                     &    50, 75, 100              &    75                                        \\ \hline
\end{tabular}%
}
\end{center}
\caption{HIPT-ABMIL hyperparameters tuned using a three-stage grid search. The same hyperparameters were tuned for ResNet-ABMIL and HistoResNet-ABMIL. An additional hyperparameter was tuned for HIPT-CLAM to set the number of regions used for clustering \cite{Lu2021}.}
\label{tab:tuning}
\end{table}

\section{Results}


\begin{table}[h]
\begin{center}
\resizebox{0.87\textwidth}{!}{%
\begin{tabular}{ | >{\centering\arraybackslash}m{0.2\textwidth} | >{\centering\arraybackslash}m{0.2\textwidth} | >{\centering\arraybackslash}m{0.2\textwidth} | >{\centering\arraybackslash}m{0.2\textwidth} | >{\centering\arraybackslash}m{0.2\textwidth} | }
\hline
\rowcolor[HTML]{707070} 
{\color[HTML]{FFFFFF} Method} & {\color[HTML]{FFFFFF} AUC}                            & {\color[HTML]{FFFFFF} Balanced Accuracy} & {\color[HTML]{FFFFFF} Accuracy}  & {\color[HTML]{FFFFFF} F1 Score} \\ \hline
\cellcolor[HTML]{707070}{\color[HTML]{FFFFFF} Baseline* \cite{Wang2022b}}                                                                                                     & NA &      NA                     & 88.2\% $\pm$ 6\%*                        & 0.917 $\pm$ 0.07*
\\  \hline
\cellcolor[HTML]{707070}{\color[HTML]{FFFFFF} HIPT-ABMIL} & 0.646 $\pm$ 0.033                                                                                                & \textbf{60.2\% $\pm$ 2.9\%}             &   \textbf{61.0\% $\pm$ 2.9\%}              & 0.656 $\pm$ 0.031                  \\ \hline
\cellcolor[HTML]{707070}{\color[HTML]{FFFFFF} HIPT-CLAM} & 0.624 $\pm$ 0.033                                                                               &        57.6\% $\pm$ 2.9\%           &     58.9\% $\pm$ 2.9\%                          & 0.650 $\pm$ 0.031                  \\ \hline
\cellcolor[HTML]{707070}{\color[HTML]{FFFFFF} ResNet-ABMIL} & 0.569 $\pm$ 0.034 & 52.7\% $\pm$ 2.9\% & 54.3\%  $\pm$ 3.0\%                              & 0.617 $\pm$ 0.031                \\ \hline
\cellcolor[HTML]{707070}{\color[HTML]{FFFFFF} HistoResNet-ABMIL} & \textbf{0.655 $\pm$ 0.032} & 58.1\% $\pm$ 2.9\% & 59.6\%  $\pm$ 2.9\%                              &  \textbf{0.660 $\pm$ 0.030}                 \\ \hline
\end{tabular}%
}
\caption{5-fold cross-validation classification performance on the internal 282 WSIs (mean $\pm$ standard deviation from 100,000 iteration bootstrapping). *The best results are highlighted in bold except for the baseline results, which were generated by different researchers using different validation methods \cite{Wang2022b}, as described in the Discussion. HIPT-ABMIL and HIPT-CLAM are the combined transformer and multiple instance learning approaches used in this paper. ResNet-ABMIL is the standard attention-based multiple instance learning \cite{Ilse2018} approach using an ImageNet-pretrained ResNet50 encoder. HistoResNet-ABMIL is the same approach using a histopathology-pretrained ResNet18 encoder \cite{Ciga2022}. HIPT - Hierarchical Image Pyramid Transformer \cite{Chen2022}. \mbox{ABMIL - Attention-based} Multiple Instance Learning \cite{Ilse2018}. \mbox{CLAM - Clustering-constrained-attention} Multiple Instance Learning \cite{Lu2021}. \mbox{AUC - Area} under the receiver operating characteristic curve.}
\label{tab:results}
\end{center}
\end{table}
\vspace{-0.5cm}

The HIPT-ABMIL model had the greatest performance for two evaluated metrics and HistoResNet-ABMIL had the greatest performance for the other two. On the internal 5-fold test set, the HIPT-ABMIL model achieved an AUC of 0.646$\pm$0.033, balanced accuracy of 60.2\%$\pm$2.9\%, and F1 score of 0.656$\pm$0.031 (mean $\pm$ one standard deviation from 100,000 iteration bootstrapping). The results were highly varied, with the AUC per cross-validation fold being \mbox{0.381-0.825}. There were also large differences between validation and test set performance in most folds, including a fold where validation AUC was 0.400 higher \mbox{(0.781 vs. 0.381)} and another where test AUC was 0.389 higher (0.436 vs. 0.825). The performance of the histopathology-pretrained models was much greater than the ImageNet-pretrained ResNet-ABMIL, which achieved just 52.7\% balanced accuracy, barely greater than random guessing. No clear classification benefit was found from using hierarchical transformers compared to a ResNet, or from using CLAM rather than standard ABMIL.  The optimal HIPT-ABMIL and HistoResNet-ABMIL models were each applied to the external ATEC23 TMA test set, though neither generalised well to this data (accuracies of 35\% and 55\% respectively). 

\begin{figure}[h]
    \centering
\includegraphics[trim={0.5cm 0.5cm 0.5cm 0.5cm},width=0.6\textwidth]{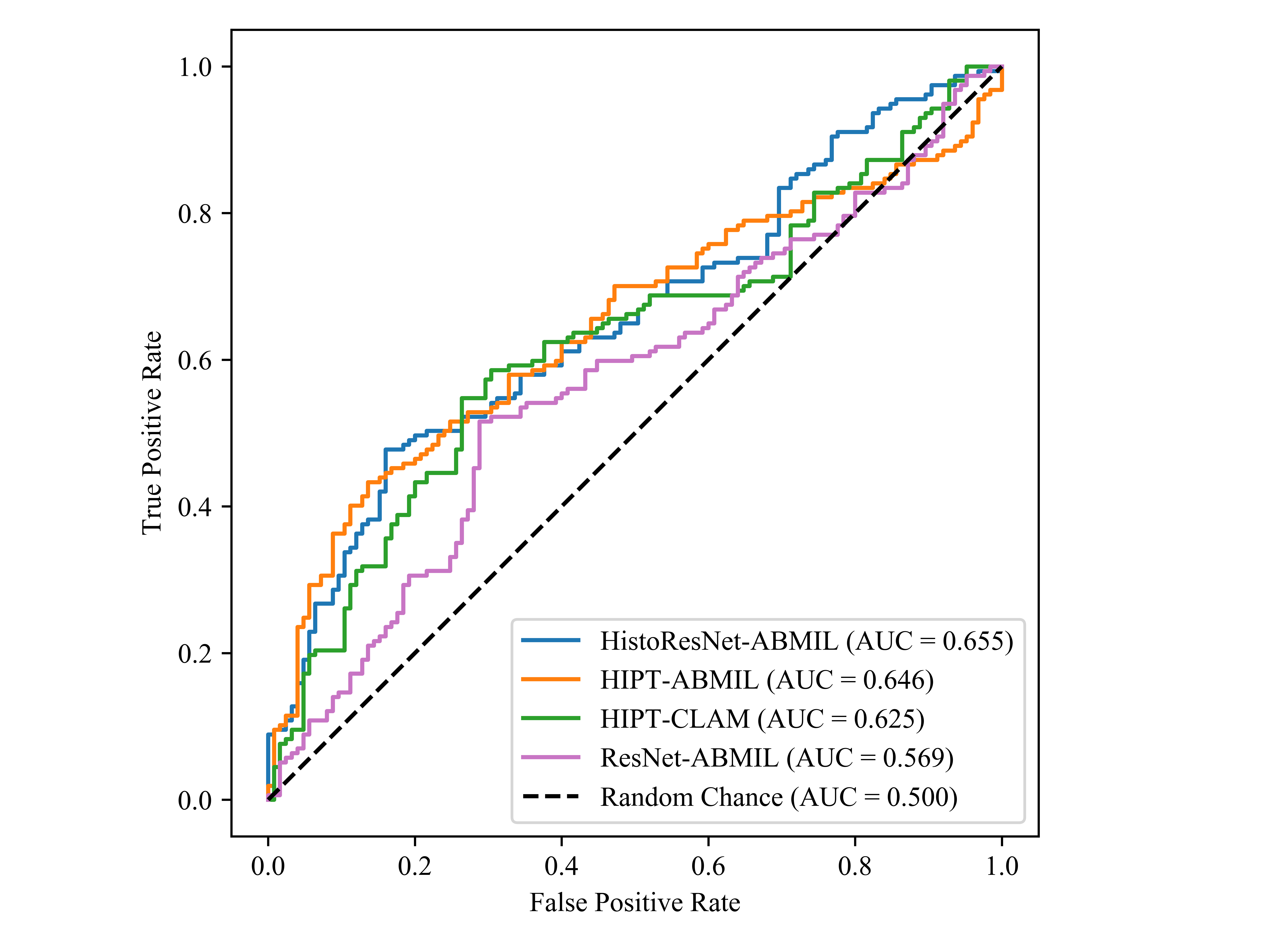}
    \caption{Receiver operating characteristic (ROC) curves and the area under curve (AUC) for each model from 5-fold cross-validation.}
    \label{plt:roc}
\end{figure} 

\section{Discussion}
Our internal performance scores were much lower than the reported performance of the baseline approach (optimal F1 of 0.660 compared to 0.917, accuracy of 60.2\% compared to 88.2\% \cite{Wang2022b}). However, this is unlikely to be a fair comparison due to differences in the pre-processing, validation, and data used. Further validation would be beneficial in evaluating both approaches as there is a high risk that results were artificially inflated by confounding and bias caused by the high levels of heterogeneity in the relatively small dataset. We partially mitigated this by splitting data into train-val-test splits per patient, reducing the unduly high level of correlation between training and testing sets. However, there were other likely confounders which were not adequately controlled, with the dataset containing small quantities of WSIs with significant differences to the majority, such as different carcinoma histological subtypes (clear cell, endometrioid, mucinous, and unclassified carcinomas), tissue types/background histology (omentum, peritoneum, lymph node) and artifacts (pen markings, image stitching, out of focus regions). Such confounding could be moderated by using a larger, more clinically representative dataset. The large standard deviations in the results were also likely attributable to the relatively small dataset size, with a 95\% confidence interval for the optimal balanced accuracy being 54.5\% to 66.0\%. No challenge participant achieved an accuracy greater than random chance, which may indicate that TMAs do not contain sufficient prognostic information.

The clinical utility of these models would benefit from a more precise and clinically relevant definition of outcome, as the ATEC23 binary classification grouped patients who relapsed after just over 6 months together with patients who never relapsed. Significant consideration should be given to the impact of carcinoma stage, grade and morphological subtype on outcome beyond the Cox models presented in previous research, which found strong but not statistically significant correlations between the subtype and outcome, and between the stage and outcome \cite{Wang2022b}. Further details about the cohort’s patients in terms of their differing responses to platinum-based chemotherapy would also be informative as the model may be predicting response to a mixture of combination and single-agent therapies.

\begin{figure}[b]
    \centering
    \begin{subfigure}[b]{0.31\textwidth}
         \centering
         \includegraphics[width=\textwidth]{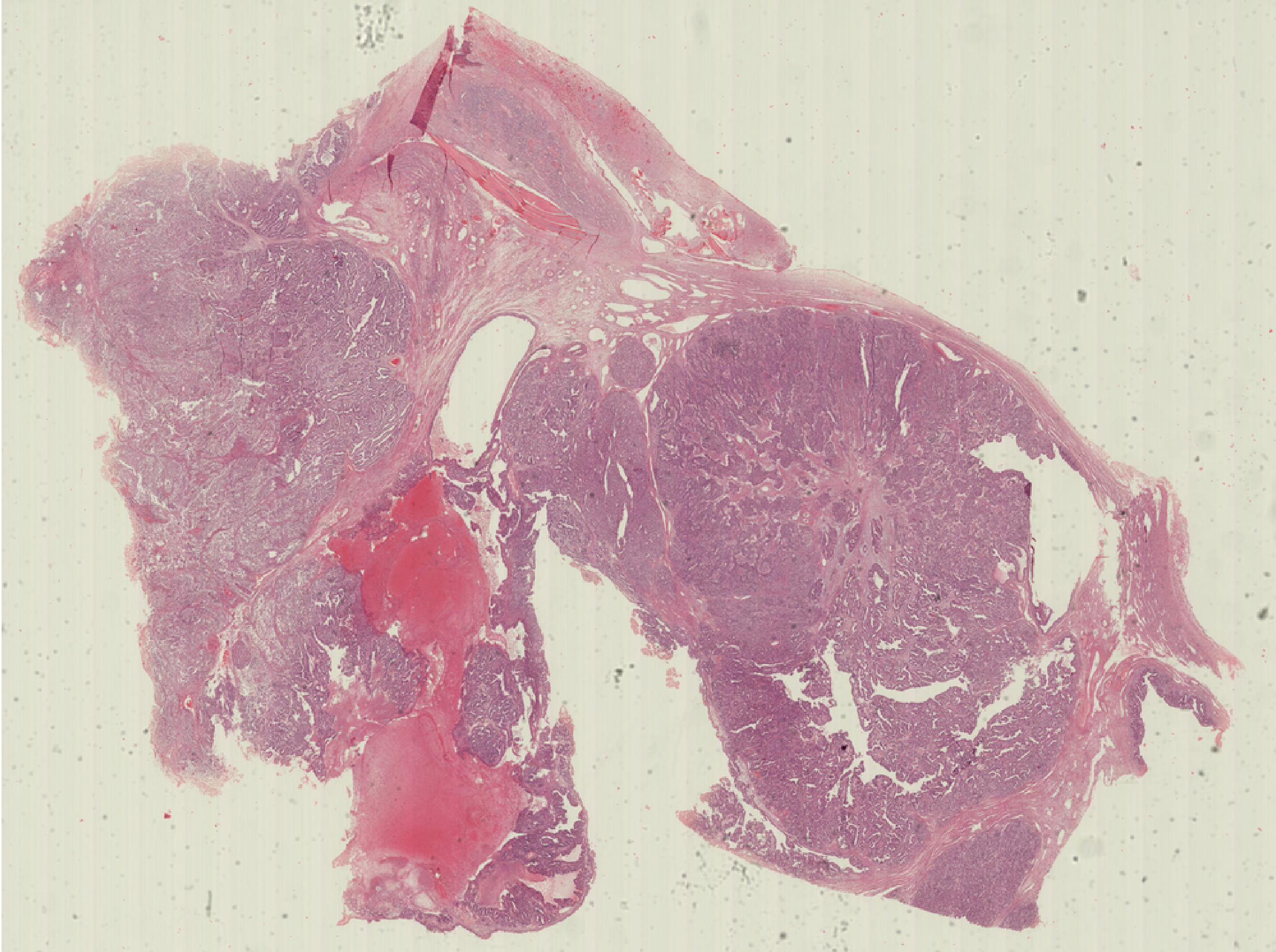}
         \caption{Raw histopathology slide from the ATEC23 challenge training set.}
         \label{fig:rawslide}
     \end{subfigure}
     \hfill
     \begin{subfigure}[b]{0.31\textwidth}
         \centering
         \includegraphics[width=\textwidth]{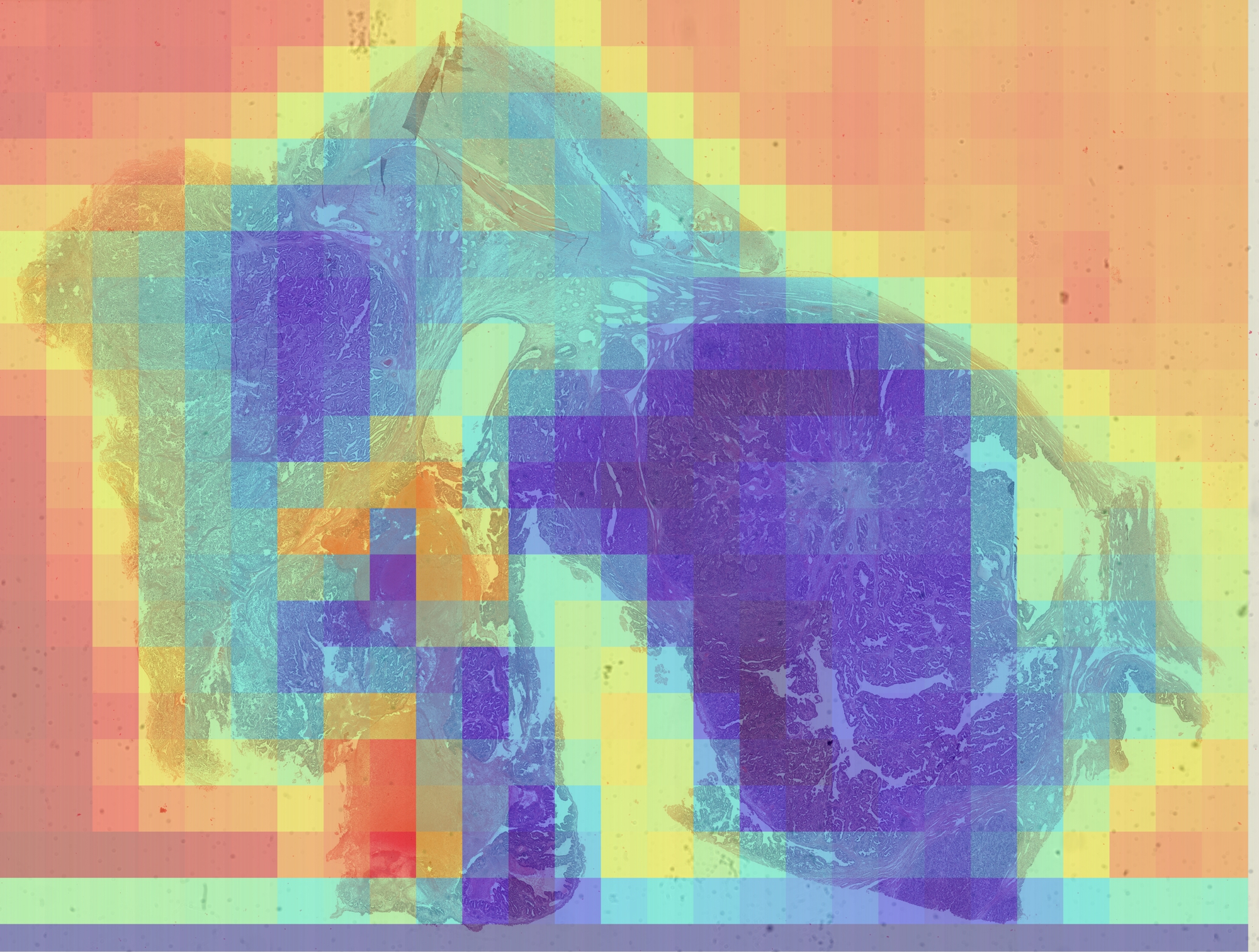}
         \caption{Corresponding ABMIL heatmap from the initial HIPT-ABMIL model.}
         \label{fig:badheatmap}
     \end{subfigure}
    \hfill
     \begin{subfigure}[b]{0.31\textwidth}
         \centering
         \includegraphics[width=\textwidth]{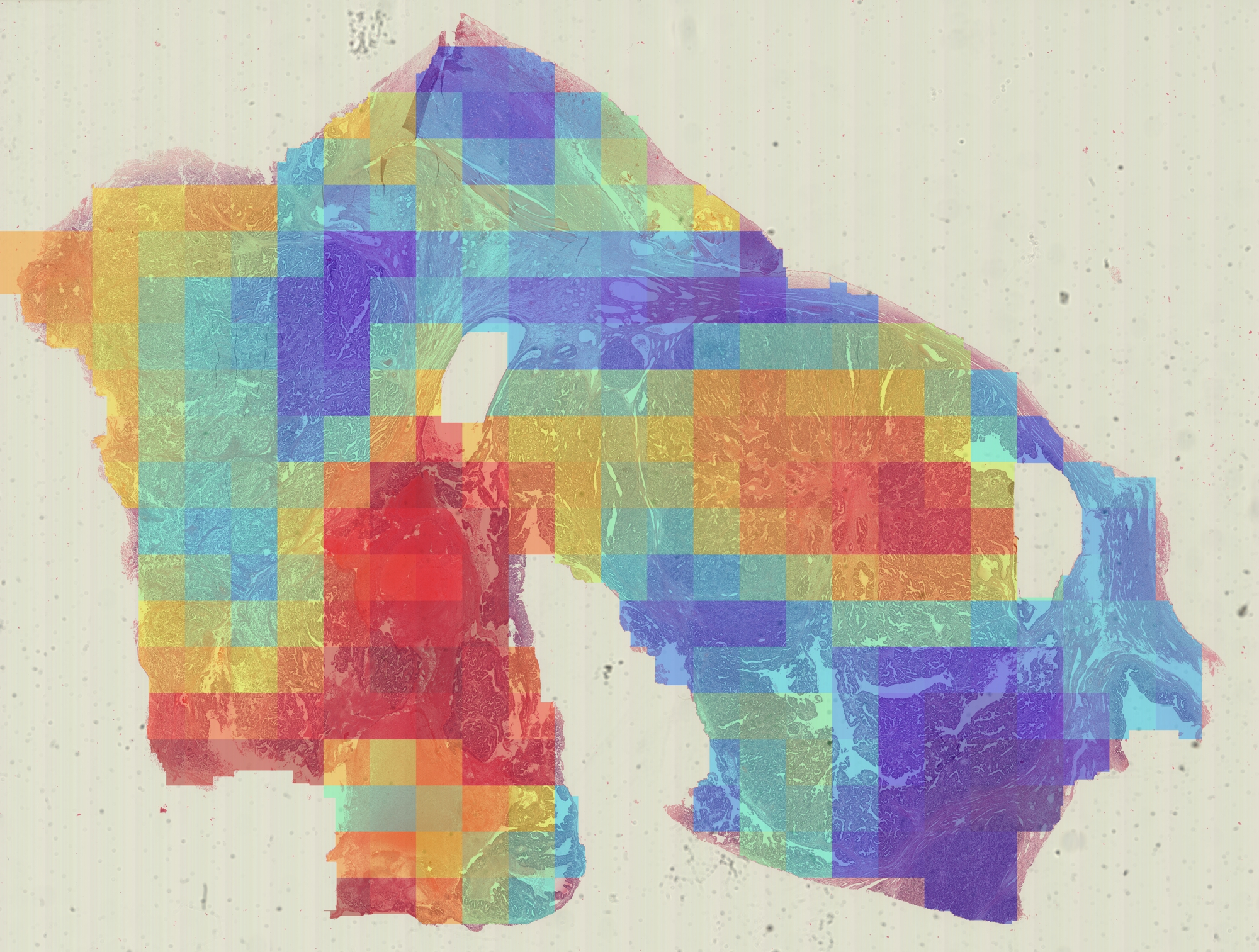}
         \caption{Corresponding ABMIL heatmap from the final HIPT-ABMIL model.}
         \label{fig:goodheatmap}
     \end{subfigure}
    \caption{Example of a WSI in which the initial model heatmap exhibited a high level of confounding, as evidenced by the background receiving greater attention than the tissue. In the final model, the tissue segmentation preprocessing step was improved to avoid including background, leading to heatmaps that do not exhibit clear confounding.}
    \label{fig:heatmaps}
\end{figure}

In our initial experiments, attention heatmaps showed that, in some WSIs, background regions were given much higher attention scores than tissue regions, indicating that these slides were being classified according to irrelevant information. The chromatic variability of WSIs was leading to inconsistent tissue segmentations, with some including a large amount of non-tissue areas. As a result, we adjusted our tissue segmentation parameters to achieve a more consistent performance, with the changes to a resulting heatmap shown in Figure \ref{fig:heatmaps}. All results presented in this paper were generated using these updated segmentations. Before this update, our 5-fold cross-validation accuracy was 89.7\% and the F1 score was 0.915, which was very similar to the reported baseline model performance \cite{Wang2022b}. Improving the initial background segmentation significantly reduced internal classification performance, indicating that the slide backgrounds contained confounding information that could artificially inflate internal performance. This highlights the need for explainability in digital pathology AI to understand any model's decision-making process.


\section{Conclusion}
It is unclear whether treatment response can be accurately predicted from ovarian cancer histopathology slides alone, with our results indicating that whole slide images may contain some prognostic signal that can be leveraged using hierarchical transformers and attention-based multiple instance learning. We found that it was beneficial to use feature extractors that were pretrained using large sets of histopathology data, though did not find transformer-based models to outperform ResNet-based models. Given that the internal experiments were conducted on a set of only 282 histopathology WSIs from 78 patients and that external validations were conducted only on TMAs, more robust validations are required before the scale of the clinical utility of these algorithms can be more reliably evaluated.

\section*{Acknowledgements}
Jack Breen is supported by the UKRI Engineering and Physical Sciences Research Council (EPSRC) [EP/S024336/1]. For the purpose of open access, the author has applied a Creative Commons Attribution (CC BY) licence to any Author Accepted Manuscript version arising from this submission.





\bibliography{articlerefs}

\end{document}